\documentclass[11pt, tightenlines, onecolumn, amsmath, amssymb, notitlepage, a4paper, nofootinbib, superscriptaddress]{revtex4-1}

\usepackage[strict]{revquantum}

\usepackage{graphicx}
\usepackage{algpseudocode}
\usepackage{braket}
\usepackage{bbm}
\usepackage{enumitem}
\usepackage[normalem]{ulem}

\usepackage{xcolor}

\usepackage{hyperref}
\hypersetup{linkcolor=blue,urlcolor=blue,citecolor=magenta,pdfstartview={FitH},hyperfootnotes=false,unicode=true}

\def\im{{\rm i}}
\def\e{{\rm e}}
\def\d{{\rm d}}

\newcommand{\Trs}[1]{\operatorname{\textnormal{Tr}}\left[ {#1} \right]}
\newcommand{\norm}[1]{\left\lVert#1\right\rVert}

\begin{document}

\title{Universally Robust Control of Open Quantum Systems}

\author{Lixiang Ding}
\affiliation{School of Physics Science and Engineering, Tongji University, Shanghai 200092, China}

\author{Jingtao Fan}
\email{Corresponding author: Jingtao Fan, Email: fanjt@sxu.edu.cn}
\affiliation{State Key Laboratory of Quantum Optics Technologies and Devices, Institute of Laser Spectroscopy, Shanxi University, Taiyuan 030006, China}
\affiliation{Collaborative Innovation Center of Extreme Optics, Shanxi University, Taiyuan 030006, China}

\author{Xingze Qiu}
\email{Corresponding author: Xingze Qiu, Email: xingze@tongji.edu.cn}
\affiliation{School of Physics Science and Engineering, Tongji University, Shanghai 200092, China}

\date{\today}

\begin{abstract}
Mitigating noise-induced decoherence is the central challenge in controlling open quantum systems.
While existing robust protocols often require precise noise models, we introduce a universal framework for noise-agnostic quantum control that achieves high-fidelity operations without prior environmental noise characterization.
This framework capitalizes on the dynamical modification of the system-environment coupling through control drives, an effect rigorously encoded in the dynamical equation.
Since the derived noise sensitivity metric remains independent of the coupling details between the system and the environment, our protocol demonstrates robustness against arbitrary Markovian noises within the first-order weak coupling approximation.
Numerical validation through quantum state transfer and gate operations reveals near-unity fidelity across diverse noise regimes, achieving orders-of-magnitude error suppression compared to target-only approaches.
This framework bridges critical gaps between theoretical control design and experimental constraints, establishing a broadly applicable strategy for high-fidelity quantum information processing across platforms such as superconducting circuits, trapped ions, and solid-state qubits.
\end{abstract}

\maketitle

\section{Introduction}

Quantum computing and quantum information processing promise revolutionary advances in computing power \cite{Bharti_2022_RMP}, cryptography \cite{Portmann_2022_RMP}, and sensing \cite{Degen_2017_RMP}.
However, the fundamental challenge of maintaining quantum coherence in the face of environmental noise remains the primary obstacle to practical implementation \cite{Suter_2016_RMP}.
At the heart of these transformative technologies lies the fundamental challenge of precise manipulation of quantum states, where the fragile coherence of quantum systems must be meticulously preserved and controlled \cite{Lidar_2025_PRXQ}.
Although effective for isolated systems, traditional quantum control strategies often become inadequate for open systems due to unmodeled system-environment interactions.
Consequently, developing robust optimal control theories for open quantum systems has emerged as a critical frontier, essential for bridging the gap between idealized theoretical models and practical implementations under noisy, real-world conditions \cite{Shimshon_2022_SciAdv, Weidner_2025_RobustQC, Gautier_2025_PRL}.

Optimal control in open quantum systems must reconcile two competing objectives: driving the system toward a target state while simultaneously suppressing unwanted environmental couplings \cite{Weidner_2025_RobustQC}.
To date, several control protocols---including dynamical decoupling \cite{Viola_1999_PRL, Ryan_2010_PRL}, decoherence-free subspaces \cite{Lidar_1998_PRL, Bacon_2000_PRL}, and filter-function optimization \cite{Viola_2014_Filter, Cerfontaine_2021_Filter}---have demonstrated significant success in this endeavor.
However, their effectiveness critically depends on prior knowledge of the mathematical structure of environmental noise \cite{Lidar_2008_PRL, Schulte-Herbrüggen_2011_Optimal, Jing_2013_PRA, Maloney_2022_PRA, Liu_2024_PRA, Zou_2025_SciAdv}, a requirement that is rarely satisfied in experimental settings due to the inherent difficulties of comprehensive noise characterization.
This gap necessitates the development of universally applicable control strategies that remain effective without detailed knowledge of system-environment coupling.

Here, we introduce a universally robust quantum control framework that substantially enhances operational fidelity in open quantum systems without requiring prior knowledge of noise characteristics.
Our approach is built upon a key physical insight: external control drives not only manipulate the target system but also dynamically reconfigure the system-environment interaction itself \cite{Dann_2018_PRA, Shimshon_2022_SciAdv, DiMeglio_2024_Quantum, Dann_2025_Quantum}.
This dual effect is rigorously captured through a time-dependent master equation formalism, where the control fields instantaneously modulate the dissipative rates.
Crucially, this description remains valid beyond adiabatic approximations and maintains thermodynamic consistency \cite{Shimshon_2022_SciAdv}, establishing a physically grounded foundation for quantum control design.
Through an analytical treatment of the master equation's evolution operator, we derive an effective noise sensitivity metric that quantifies environmental susceptibility.
This metric is then incorporated into a minimized objective functional defined exclusively by task-specific system observables, creating an inherently noise-agnostic optimization landscape.
The resulting formulation provides robustness against arbitrary Markovian noises within the first-order weak coupling approximation without requiring microscopic noise models---a mathematical manifestation of our protocol's universality.
We demonstrate the efficacy of our protocol through simulations of quantum state transfer and gate operations, optimized using the Chopped Random Basis (CRAB) algorithm \cite{Doria_2011_PRL_CRB, Caneva_2011_CRB, Muller_2022_CRB}.
Both implementations achieve significant performance gains, obtaining greater than 98\% fidelity across a broad range of system-bath coupling strengths.
Our framework's versatile nature makes it broadly applicable to leading quantum hardware, including superconducting circuits \cite{Abdelhafez_2020_PRA, Werninghaus_2021_npjQI}, trapped ions \cite{Nebendahl_2009_PRA, Shapira_2018_PRL}, and nitrogen-vacancy centers \cite{Poggiali_2018_PRX,Vetter_2024_npjQI}.
This unified approach establishes a concrete approach for robust quantum control, effectively bridging the divide between theoretical robustness and experimental feasibility.

\section{Results}

\subsection{Model}

The complete quantum description of a control task is encoded in the composite Hamiltonian
\begin{equation}
\hat{H}(t)=\hat{H}_{\rm S}(t)+\hat{H}_{\rm B}+\hat{H}_{\rm I}  \, ,
\label{H_tot}
\end{equation}
where $\hat{H}_{\rm S}(t) = \hat{H}_0 + \sum_i u_i(t)\hat{H}_i$ represents the noise-free system Hamiltonian, with $\hat{H}_0$ denoting the bare drift Hamiltonian, $\hat{H}_i$ the $i$-th control operator, and $u_i(t)$ the corresponding time-dependent control field.
The bath Hamiltonian $\hat{H}_{\rm B}$ governs the environment's free dynamics, which couples to the system via interaction Hamiltonian $\hat{H}_{\rm I}=g\sum_{\alpha }\hat{A}_{\alpha }\otimes \hat{B}_{\alpha }$ (the most general form), where $\hat{A}_{\alpha }$ and $\hat{B}_{\alpha }$ are Hermitian operators of the system and the bath, respectively, while $g$ characterizes their coupling strength.
Under the Born-Markov approximation, tracing over the bath degrees of freedom yields the Gorini-Kossakowski-Lindblad-Sudarshan (GKLS) master equation \cite{Lindblad_1975_CP, Gorini_1976_CP, Lindblad_1976_On,Pearle_2012_Lindblad1,Manzano_2020_Lindblad2} (in units where $\hbar = 1$)
\begin{equation}
\frac{\d}{\d t}\hat{\rho}_{\rm S}(t)=- \im[\hat{H}_{\rm S}(t),\hat{\rho}_{\rm S}(t)]+\mathcal{D}[\hat{\rho}_{\rm S}(t)]  \,,
\label{Eq:GKLS}
\end{equation}
where $\hat{\rho}_{\rm S}(t)$ is the density operator of the system.
The dissipative superoperator $\mathcal{D}$ takes the form
\begin{equation}
\begin{aligned}
\mathcal{D}[\hat{\rho}_{\rm S}(t)]=\lambda \sum_{j}\kappa _{j}(t)
\left[\hat{F}_{j}(t)\hat{\rho}_{\rm S}(t)\hat{F}_{j}^{\dagger}(t)-\frac{1}{2}\{\hat{F}_{j}^{\dagger}(t)\hat{F}_{j}(t),\hat{\rho}_{\rm S}(t)\}\right]\, ,
\end{aligned}
\label{Eq:D}
\end{equation}
with $\lambda =g^{2}$.
The Lindblad jump operator $\hat{F}_{j}(t)$ constitute the eigenoperators of the free dynamical map $\mathcal{U}_{\rm S}(t)$, satisfying
\begin{equation}
\mathcal{U}_{\rm S}(t)\hat{F}_{j}(t)=\hat{U}_{\rm S}^{\dagger }(t)\hat{F}_{j}(t)
\hat{U}_{\rm S}(t)=\e^{\im \phi _{j}(t)}\hat{F}_{j}(t) \, ,
\label{Eq:F}
\end{equation}
where $\hat{U}_{\rm S}(t)$ is the system propagator evolving under $\im  \partial _{t}\hat{U}_{\rm S}(t)=\hat{H}_{\rm S}(t)\hat{U}
_{\rm S}(t)$ with initial condition $\hat{U}_{\rm S}(0)=\hat{\mathbb{I}}_{\rm S}$ (the identity operator for the system Hilbert space). 
With the definition Eq.~\eqref{Eq:F}, the jump operator $\hat{F}_{j}(t)$ and its corresponding phase $\phi _{j}(t)$\ can be conveniently constructed as $\hat{F}_{j}(t)=\ket{ u_{n}(t)} \!\bra{u_{m}(t)}$ and $\phi _{j}(t)=\varepsilon _{n}(t)-\varepsilon_{m}(t)$ with $j=N(n-1)+m$. Here, $N$ is the dimension of system's Hilbert space, and $\ket{u_{n}(t)} $ and $\varepsilon _{n}(t)$\ are, respectively, instantaneous eigenstates and eigenphases of $\hat{U}_{\mathrm{S}}(t)$ satisfying $\hat{U}_{\mathrm{S}}(t)\ket{ u_{n}(t)} =\exp [-i\varepsilon _{n}(t)]\ket{ u_{n}(t)}$ \cite{Shimshon_2022_SciAdv,Dann_2018_PRA,Wu_2022_PRA}.
The time-dependent decoherence rates $\kappa _{j}(t)$ read
\begin{equation}
\kappa _{j}(t)=\sum_{\alpha }\left[\eta _{j}^{\alpha } (t)\right]^{2}\gamma
_{\alpha \alpha }\left[\omega _{j}(t)\right] \, ,
\label{Eq:kappa}
\end{equation}
where $\eta _{j}^{\alpha }(t)=\left\vert \Trs{\hat{F}_{j}^{\dagger }(t)\hat{A}_{\alpha }}\right\vert $ and $\gamma _{\alpha \alpha}[\omega _{j}(t) ]=\int_{-\infty }^{\infty }\d s\,\e^{ - \im\omega_{j}(t) s}\,
{\rm Tr}_{\rm B}\left[\tilde{B}_{\alpha}(s)\tilde{B}_{\alpha}(0)\hat{\rho}_{\rm B}\right]$ with $\hat{\rho}_{\rm B}$ the density operator of the bath (see Method A for a detailed discussion on $\gamma _{\alpha \alpha}[\omega _{j}(t) ]$).
Here, $\tilde{B}_{\alpha }(s)=\hat{U}_{\rm B}^{\dagger }(s)\hat{B}_{\alpha }\hat{U}_{B}(s)$ are the bath operators in the interaction picture and $\omega _{j}(t)\left[=\d\phi _{j}(t)/\d t\right]$ account for the effective instantaneous Bohr frequencies.

Critically, the time-dependent master equation explicitly accounts for how control drives in $\hat{H}_{\rm S}(t)$ modulate the dissipative dynamics.
This results in control-dependent jump operators $\hat{F}_{j}(t)$ and time-varying decoherence rates $\kappa_{j}(t)$ \cite{Shimshon_2022_SciAdv}.
Consequently, the controller influences the system through two distinct channels: (i) Direct unitary steering via the coherent evolution $-\im[\hat{H}_{\rm S}(t),\hat{\rho}_{\rm S}(t)]$ and (ii) Indirect dissipative control through the modulated dissipator $\mathcal{D}[\hat{\rho}_{\rm S}(t)]$.
Remarkably, this framework enables the discovery of optimal controls $\hat{H}_{\rm S}(t)$ that simultaneously achieve target operations while actively suppressing environment-induced noise.
As demonstrated in the subsequent section, such robust control protocols can be identified—even without detailed knowledge of system-bath couplings—through appropriate construction of the objective functional.

\subsection{Universally Robust Noise Mitigation}

We now formulate a universal approach to suppress arbitrary Markovian noise by constructing a control-dependent cost function based on the time-dependent master equation. This cost function depends solely on the designed control Hamiltonian $\hat{H}_{\rm S}(t)$ while remaining independent of specific noise channels.

For the sake of convenience, we first vectorize the master equation Eq.
\eqref{Eq:GKLS} by rewriting it in the Hilbert-Schmidt space \cite{Caves_1999_QE, Erika_2007_FK}. 
This can be done by reshaping the density matrix $\hat{\rho}_{\rm S}(t)$ as a column vector, denoted as $|\hat{\rho}_{\rm S}(t)\rangle\!\rangle $.
It is straightforward to show that $|\hat{\rho}_{\rm S}(t)\rangle\!\rangle $ satisfies the following Schr\"odinger-type equation: 
\begin{equation}
\frac{\partial |\hat{\rho}_{\rm S}(t)\rangle\!\rangle }{\partial t}=\vec{\mathcal{L}}
(t)|\hat{\rho}_{\rm S}(t)\rangle\!\rangle \,,
\label{Eq:HS}
\end{equation}
where the Lindbladian superoperator is vectorized as
\begin{equation}
\begin{aligned}
\vec{\mathcal{L}}(t)=
&-\im\left[\hat{H}_{\rm S}(t)\otimes\hat{\mathbb{I}} - \hat{\mathbb{I}}\otimes\hat{H}_{\rm S}^{\rm T}(t)\right] \\
&+ \lambda \sum_{j}
\kappa_j(t)
\left(
\hat{F}_{j}(t)\otimes\hat{F}_{j}^{\ast }(t)
-\frac{1}{2}\left[\hat{F}_{j}^{\dagger}(t)\hat{F}_{j}(t)\otimes\hat{\mathbb{I}} + \hat{\mathbb{I}}\otimes(\hat{F}_{j}^{\dagger}(t)\hat{F}_{j}(t))^{\rm T}\right]\right) \,.
\end{aligned}
\label{Eq:L}
\end{equation}
This Hilbert-Schmidt vectorization is mathematically equivalent to the Bloch representation but is employed here as it transforms abstract superoperators into concrete $N^2 \times N^2$ matrices, which is computationally convenient for perturbation analysis and numerical optimization we applied subsequently.

By defining the shorthand $\mathbf{H}(t)=-\im[\hat{H}_{\rm S}(t)\otimes\hat{\mathbb{I}} - \hat{\mathbb{I}}\otimes\hat{H}_{\rm S}^{\rm T}(t)]$
and $\mathbf{F}^{j}(t)= \hat{F}_{j}(t)\otimes \hat{F}_{j}^{\ast }(t)
 - \frac{1}{2}[\hat{F}_{j}^{\dagger}(t)\hat{F}_{j}(t)\otimes\hat{\mathbb{I}} + \hat{\mathbb{I}}\otimes(\hat{F}_{j}^{\dagger}(t)\hat{F}_{j}(t))^{\rm T}]$,
the formal solution of Eq.~\eqref{Eq:HS} can be expressed as $|\hat{\rho}_{\rm S}(t)\rangle\!\rangle = \mathbf{V}(t)|\hat{\rho}(0)\rangle\!\rangle$.
Here,
\begin{equation}
\begin{aligned}
\mathbf{V}(t)=\mathcal{T}\exp \left(\int_{0}^{t}\d t_{1}\left[\mathbf{H}(t_{1})+
\lambda \sum_{j}\kappa_j(t_1) \mathbf{F}^{j}(t_{1})\right]\right)
\end{aligned}
\end{equation}
is the time-evolution operator with $\mathcal{T}$ the chronological time-ordering operator.
To isolate the noise effects, we transform into the interaction picture with respect to the noise-free Hamiltonian $\hat{H}_{\rm S}(t)$ and separate the evolution operator into a unitary part and an error part, 
\begin{equation}
\mathbf{V}(t)=\mathbf{U}(t)\mathbf{U}_{\mathrm{err}}(t)\, ,
\end{equation}
where $\mathbf{U}(t)=\mathcal{T}\exp \left[ \int_{0}^{t}\d t_{1}\mathbf{H}(t_{1})\right]$ accounts for unitary evolution and any noise-induced non-unitary effects are attributed to the operator $\mathbf{U}_{\mathrm{err}}(t)=\mathcal{T}\exp \left[ \lambda \int_{0}^{t}\d t_{1}  \sum_{j}\kappa_j(t_1)\widetilde{\mathbf{F}}^{j}(t_{1})\right] $ with $\widetilde{\mathbf{F}}^{j}(t)=\mathbf{U}^{\dagger }(t)\mathbf{F}^{j}(t)\mathbf{U}(t)$.

Under the assumption of weak system-bath coupling, we can expand the
evolution operator $\mathbf{U}_{\mathrm{err}}(t)$ in terms of $\lambda $,
leading to
\begin{equation}
\begin{aligned}
\mathbf{U}_{\mathrm{err}}(t) = \hat{\mathbb{I}}_{\rm S}\otimes  \hat{\mathbb{I}}_{\rm S}
&+\lambda \int_{0}^{t}\d t_{1}\sum_{j}\kappa_{j}(t_1)\widetilde{\mathbf{F}}^{j}(t_{1})   \\
&+\lambda^{2}\int_{0}^{t}\d t_{1}\int_{0}^{t_{1}}\d t_{2}\sum_{j_{1},\,j_{2}}
\kappa_{j_{1}}(t_{1})\kappa _{j_{2}}(t_{2})\widetilde{\mathbf{F}}^{j_{1}}(t_{1})\widetilde{\mathbf{F}}^{j_{2}}(t_{2})+\cdots \,.
\end{aligned}
\label{Eq:Uerr}
\end{equation}
Robust quantum control requires insensitivity to first-order effects in the system-bath coupling strength $\lambda$ \cite{Note1}.
We therefore focus on the leading-order contribution, namely,
\begin{eqnarray}
\left. \frac{\d\mathbf{U}_{\mathrm{err}}(t)}{\d\lambda }\right\vert _{\lambda=0}
= \int_{0}^{t}\d t_{1}\sum_{j}\kappa _{j}(t_{1})\widetilde{\mathbf{F}}^{j}(t_{1})
= \int_{0}^{t}\d t_{1}\sum_{\alpha,\, j }\left[\eta _{j}^{\alpha } (t_1)\right]^{2}\gamma _{\alpha \alpha }\left[\omega _{j}(t_{1})\right]\widetilde{\mathbf{F}}^{j}(t_{1})\,.
\label{Eq:dUerr}
\end{eqnarray}
The norm of Eq.~\eqref{Eq:dUerr} then quantifies the noise sensitivity, the minimization of which amounts to increasing the robustness of a control protocol.
Aiming at acquiring a universal robustness, we first apply the Cauchy-Schwarz inequality to $[\eta _{j}^{\alpha } (t_1)]^{2}$, yielding
\begin{eqnarray}
\left[\eta _{j}^{\alpha } (t_1)\right]^{2}  = \left\vert \Trs{\hat{F}_{j}^{\dagger}(t_1)\hat{A}_{\alpha }}\right\vert ^{2}
\leq \Trs{ \hat{F}_{j}^{\dagger }(t_1)\hat{F}_{j}(t_1)}
\cdot \Trs{\hat{A}_{\alpha }^{\dagger }\hat{A}_{\alpha }}
=\Trs{\hat{A}_{\alpha }^{\dagger }\hat{A}_{\alpha }}\,.
\label{Eq:eta}
\end{eqnarray}
Substituting Eq.~\eqref{Eq:eta} into Eq.~\eqref{Eq:dUerr}, we have
\begin{equation}
\begin{aligned}
\norm{ \left. \frac{\d\mathbf{U}_{\mathrm{err}}(t)}{\d\lambda }\right\vert_{\lambda =0}}
\leq \sum_{\alpha }\Trs{\hat{A}_{\alpha}^{\dagger }\hat{A}_{\alpha }}\cdot \norm{\mathbf{F}_{\alpha}}
\leq \sqrt{\sum_{\alpha } \Tr^2\left[\hat{A}_{\alpha }^{\dagger }\hat{A}_{\alpha }\right]}
\cdot \sqrt{\sum_{\alpha }\norm{ \mathbf{F}_{\alpha}} ^{2}}\, ,
\end{aligned}
\label{Eq:Universal}
\end{equation}
where $\mathbf{F}_{\alpha }=\int_{0}^{t}\d t_{1}\sum_{j}\gamma _{\alpha \alpha
}\left[\omega _{j}(t_{1})\right] \widetilde{\mathbf{F}}^{j}(t_{1})$ and $\norm{\mathbf{F}_\alpha} = \sqrt{\text{Tr}[\mathbf{F}^{\dagger}_\alpha\mathbf{F}_\alpha]}$ denotes the Frobenius norm. 
In deriving the last inequality in Eq. \eqref{Eq:Universal}, we have made use of the Cauchy-Schwarz inequality twice.
Equation \eqref{Eq:Universal} suggests an effective noise sensitivity $D_{\rm eff}$, defined as (see Method B for a detailed recipe to obtain $D_{\rm eff}$)
\begin{equation}
D_{\mathrm{eff}}=\sqrt{\sum_{\alpha }\norm{\mathbf{F}_{\alpha}} ^{2}}\,.
\label{Eq:Deff}
\end{equation}
It can be seen clearly that the goal of increasing the robustness of a control problem comes down to finding a $\hat{H}_{\rm S}(t)$ to minimize $D_{\rm eff}$.
We emphasize that this noise mitigation scenario is quite universal in the sense that the effective noise sensitivity $D_{\rm eff}$ is independent of the specific system-bath coupling operators $\hat{A}_{\alpha }$ and is therefore effective in all Markovian environments.
Moreover, while we here focus on the first-order contribution of the noise perturbation, higher-order expansion terms from Eq.
(\ref{Eq:Uerr}) could in principle be incorporated into $D_{\rm eff}$ to further enhance robustness.

\subsection{Optimal Control Framework}

We now demonstrate how our universal robustness measure $D_{\rm eff}$ integrates with quantum optimal control.
The primary objective is to implement a target unitary operation $\hat{U}_{\rm tar}$ by optimizing the control fields $u_i(t)$ in the system Hamiltonian $\hat{H}_{\rm S}(t)$ [see details below Eq.~\eqref{H_tot}].
This is achieved by maximizing fidelity between $\hat{U}_{\rm tar}$ and the realized evolution $\hat{U}_{\rm S}(\tau)$, where $\tau$ is total evolution time.
We distinguish two control paradigms:
(i) For the task of state transfer from an initial pure state $\hat{\rho} _{\rm i}$ to a target pure state $\hat{\rho} _{\rm tar}$, such fidelity reduces to
\begin{equation}
\mathcal{F}_{\text{state}}=\Trs{\hat{\rho} _{\rm f} \hat{\rho} _{\rm tar}} \,.
\end{equation}
where $\hat{\rho} _{\rm f}= \hat{U}_{\rm S} (\tau)\hat{\rho} _{\rm i}\hat{U}^{\dagger }_{\rm S} (\tau)$ and $\hat{\rho} _{\rm tar}= \hat{U}_{\rm tar} \hat{\rho} _{\rm i}\hat{U}^{\dag }_{\rm tar} $.
(ii) For quantum gate operations, on the other hand, the fidelity is formulated in terms of a complete set of pure initial states $\{\hat{\rho} _{\rm i}^{n}\}$ as
\begin{equation}
\mathcal{F}_{\text{gate}}=\frac{1}{N^{2} - 1}\sum_{n=1}^{N^{2} - 1}
\Trs{ \hat{\rho} _{\rm f}^{n} \hat{\rho} _{\rm tar}^{n} } \, ,
\end{equation}
where $\hat{\rho} _{\rm f}^{n}= \hat{U}_{\rm S}(\tau)\hat{\rho} _{\rm i}^{n}\hat{U}^{\dagger }_{\rm S}(\tau) $, $\hat{\rho}_{\rm tar}^{n}= \hat{U}_{\rm tar} \hat{\rho} _{\rm i}^{n}\hat{U}^{\dag }_{\rm tar} $. 
Note that we have excluded the identity since it is preserved in the CPTP map and this definition guarantees $\mathcal{F}_{\text{gate}}\leq 1$.
In the following, we optimize the \textit{infidelity} $\mathcal{J}_0 = 1 - \mathcal{F}_{\rm state/gate}$ for numerical efficiency, converting fidelity maximization to minimization of a positive definite functional.

Robust optimal control additionally requires that the control process is immune environment noise.
We then face a multi-objective optimization task in which the total objective functional to be minimized can be organized as \cite{Poggi_2024_PRL}
\begin{equation}
\mathcal{J}=\mathcal{J}_{0}+cD_{\rm eff}\, ,
\label{Eq:J}
\end{equation}
where $c$ is a weighting coefficient that balances operational accuracy ($\mathcal{J}_{0}$) against noise robustness ($D_{\rm eff}$).
This leads to two distinct control strategies: 
\begin{itemize}
    \item \textit{Target-only control} ($c = 0$): Optimizes for fidelity alone.
\item \textit{Universally robust control} ($c > 0$): Jointly optimizes for fidelity and noise suppression.
\end{itemize}
We implement both control strategies using the CRAB algorithm \cite{Doria_2011_PRL_CRB, Caneva_2011_CRB, Muller_2022_CRB}, which offers efficient optimization with experimentally advantageous features like inherent pulse smoothness and a reduced parameter space.
The specific parameterization is $ u_i(t) = \exp( -[(t - \tau/2)/(2 \sigma ) ]^2 ) \sum_{k=1}^M c_k \sin(\nu_k t) $, where ${c_k}$ are optimization coefficients, $\sigma$ controls the Gaussian envelope width, $\tau$ denotes the total control time, and $ \nu_k=k\pi/\tau$ ($k=1,\dots,M$) are fixed frequencies (in the following, we choose $\sigma = \tau/4$ and $M=10$).
We optimize ${c_k}$ using a quasi-Newton algorithm to minimize $\mathcal{J}$.

\subsection{Performance Evaluation of Universally Robust Control}

To validate our framework, we implement the universally robust control protocol for two fundamental quantum tasks: (i) State transfer and (ii) Quantum gate operations under environmental noise.
In the simulations of these control tasks, we model the environment as a thermal bath of harmonic oscillators with $\hat{H}_{\rm B}=\sum_{k}\left( \nu _{k}\hat{a}_{k}^{\dagger }\hat{a}_{k}+\frac{1}{2}\right)$, where $\hat{a}_{k}$ are bosonic annihilation operators for mode $k$ \cite{Breuer_2002_OpenQS}.
The system-environment coupling has the generic form $\hat{H}_{\rm I}=g\hat{A}\otimes \hat{B}$, where $\hat{B}=\sum_{k}\left( g_{k}/g\right) (\hat{a}_{k}+\hat{a}_{k}^{\dagger })$ and a super-Ohmic spectral density $J(\omega )=\omega_{\rm c}^{-2}\omega ^{3}\e^{-\omega /\omega _{\rm c}}$ is assumed \cite{Weiss_2021_Dissipative}.
Here, $\omega_{\rm c}$ is the cutoff frequency, chosen as 10 times of the typical Bohr frequency of the bare system.
Such a parameter choice corresponds to a large class of quantum systems \cite{Alkauskas_2014_NV, Jahnke_2015_SV, Norambuena_2016_PRB}.

\begin{figure}[htp!]
\centering
\includegraphics[width=1.0\textwidth]{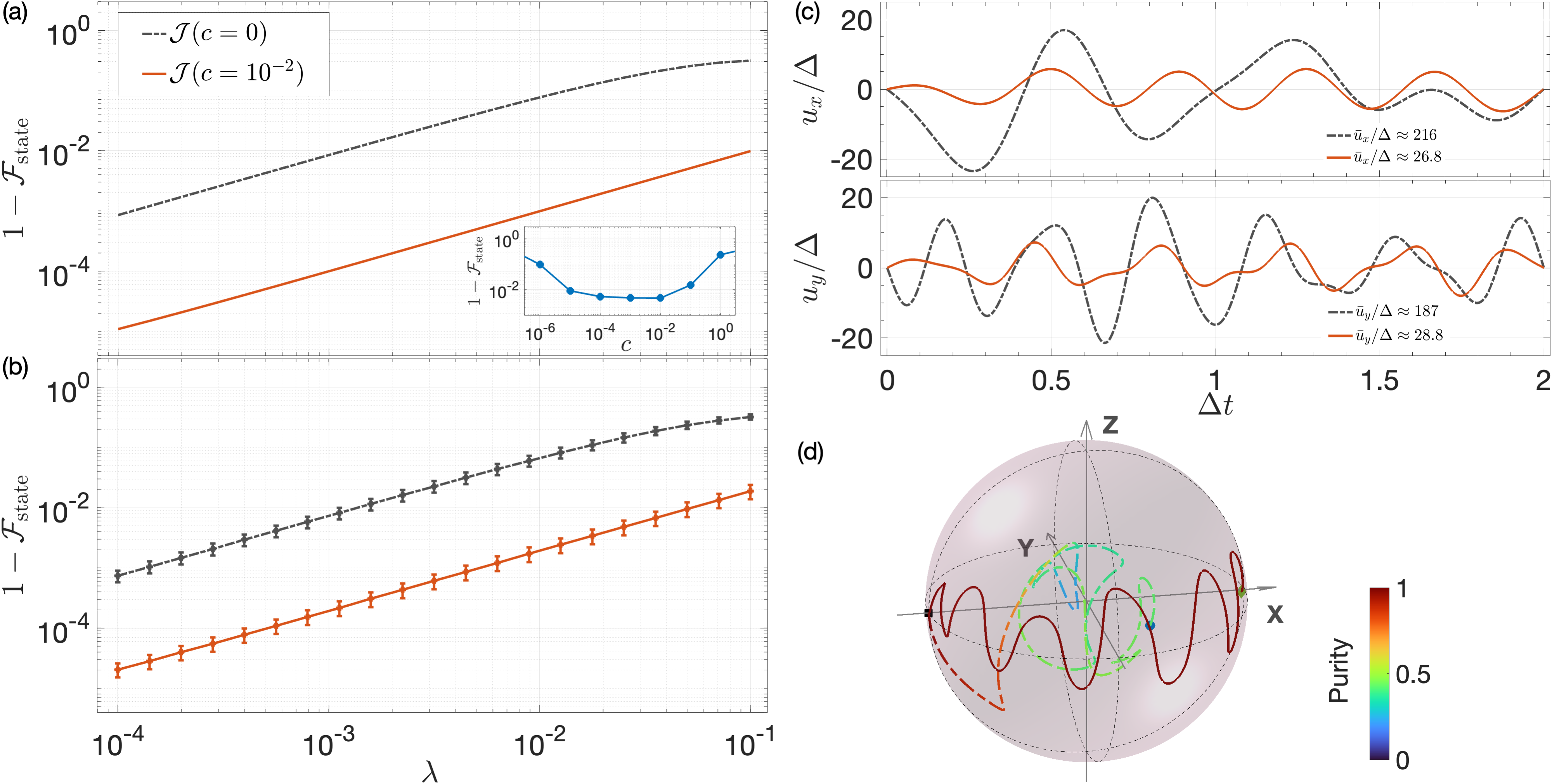}
\caption{
Universally robust quantum state transfer in a two-level system.
Quantum state transfer from initial state $\hat{\rho}_{-x}$ to target state $\hat{\rho}_{+x}$ under environmental noise. Dash-dotted lines: target-only control ($c=0$); solid lines: universally robust control ($c=10^{-2}$).
(a) State infidelity $1-\mathcal{F}_{\rm state}$ versus coupling strength $\lambda$ for specific noise ($\hat{A} = \hat{\sigma}_z$). Inset: Infidelity versus $c$ for $\lambda=0.1$.
(b) $1-\mathcal{F}_{\rm state}$ for generic noise ($\hat{A} = \mathbf{n}\cdot\hat{\boldsymbol{\sigma}}$; $\mathbf{n}$: random unit vector). 
Data averaged over 100 realizations and the error bars indicate the standard deviation. 
(c) Optimized control fields $u_x(t)$ and $u_y(t)$ for $\hat{A} = \hat{\sigma}_z$ and $\lambda=0.1$. The values of the time-integrated control power, $\bar{u}_{x,y}=\int_{0}^{\tau }u_{x,y}^{2}(t){\rm d}t$, are indicated in the legends.
(d) State evolution trajectories on Bloch sphere for $\hat{A} = \hat{\sigma}_z$ and $\lambda=0.1$. The evolution trajectory is color-coded to represent the purity of the state. Initial state: black square. Final states: blue dot (target-only) and green diamond (universally robust control).
Parameters: Total control time $\tau=2/\Delta$ and inverse temperature $\beta = 1/\Delta$.
}
\label{Fig:ST}
\end{figure}

\subsubsection{State Transfer in a Two-Level System}

We first demonstrate our quantum control protocol for state transfer implementation in a driven two-level system governed by the Hamiltonian \cite{Liu_2024_PRA}:
\begin{equation}
\hat{H}_{\rm S}(t)=\frac{\Delta }{2}\hat{\sigma} _{z}+\frac{u_{x}(t)}{2}\hat{\sigma}  _{x}+\frac{u_{y}(t)}{2}\hat{\sigma}  _{y} \, ,
\label{Eq:H_ST}
\end{equation}
where $\hat{\sigma}_{i}$ ($i=x,y,z$) denote Pauli matrices and $u_{x,\,y}(t)$ correspond to independent control fields. 
Hereafter, we set the detuning $\Delta$ as the energy unit. 
This system exhibits complete controllability as the drift Hamiltonian ($\hat{\sigma}_z$) and the control Hamiltonians ($\hat{\sigma}_{x,\,y}$) generate the full $\mathfrak{su}(2)$ Lie algebra \cite{Huang_1983_Controllability, D_Alessandro_2022}. 
Note that our chosen model with $\Delta \neq 0$ is the standard and most experimentally relevant one \cite{Chen_2016_PRL, Kuzmanovi_2024_PRR}.

We implement quantum state transfer between orthogonal maximally coherent states: initial state $\hat{\rho}_{-x} = \frac{1}{2}(\hat{\mathbb{I}} - \hat{\sigma}_x)$ and target state $\hat{\rho}_{+x} = \frac{1}{2}(\hat{\mathbb{I}} + \hat{\sigma}_x)$.
To evaluate protocol robustness, we consider two distinct noise channels: (i) specific coupling $\hat{A} = \hat{\sigma}_z$ and (ii) generic stochastic coupling $\hat{A} = \mathbf{n} \cdot \hat{\boldsymbol{\sigma}}$ with $\mathbf{n}$ being a randomly oriented unit vector.
Control fields were optimized by minimizing the objective functional $\mathcal{J}$ [Eq.~\eqref{Eq:J}] via the CRAB algorithm, comparing target-only ($c=0$) versus robust ($c>0$) control strategies.

Fig.~\ref{Fig:ST} highlights the striking advantages of our universally robust protocol. 
For a specific noise channel $\hat{A} = \hat{\sigma}_z$, while the infidelities for both protocols present an increasing tendency as $\lambda$ increases, their performance in completing the control task is distinctly different.
Under increasing $\lambda$, the target-only optimization exhibits rapid fidelity degradation ($1-\mathcal{F}_{\text{state}} > 0.3$ at $\lambda=0.1$), while the robust protocol maintains $1-\mathcal{F}_{\text{state}} < 0.01$, achieving an infidelity two orders of magnitude lower than the target-only case, throughout $\lambda \in [10^{-4},10^{-1}]$. Note that in the true $\lambda=0$ limit, the target-only control ($c=0$) can yield a fidelity equal to or better than the robust control ($c>0$), as it optimizes only for noiseless infidelity $\mathcal{J}_0$. 
Besides, while a finite $c$ should definitely improve the robustness of the control against noise, it may still damage the final fidelity at some relatively large values due to the subtle trade-off between fidelity and robustness [see  Eq.~\eqref{Eq:J}]. This trade-off is clearly demonstrated in the inset of Fig.~\ref{Fig:ST} (a), which reveals a broad ``sweet spot'' for $c$ (roughly in 
the range $10^{-4}$ to $10^{-1}$) where one can achieve a massive reduction in noise sensitivity.

When faced with generic, uncharacterized noise $\hat{A} = \mathbf{n} \cdot \hat{\boldsymbol{\sigma}}$ [Fig.~\ref{Fig:ST} (b)], the universally robust protocol maintains consistently high fidelity across all 100 random realizations ($1-\mathcal{F}_{\text{state}} < 0.02$ at $\lambda=0.1$), whereas the target-only approach fails catastrophically ($1-\mathcal{F}_{\text{state}} > 0.3$). 
This clearly confirms the universality of our robust control protocol. 
Beyond its superior fidelity, the robust protocol is also significantly more efficient, reducing the required control field amplitudes by more than half compared to the target-only method [Fig.~\ref{Fig:ST} (c)]. 
This efficiency is also reflected in the time-integrated control power (fluence), $\bar{u}_{x,y}=\int_{0}^{\tau }u_{x,y}^{2}(t){\rm d}t$, which is significantly lower for the robust protocol as shown in the figure's legend. 
Furthermore, it excels at preserving quantum state purity. As illustrated by the Bloch sphere trajectories which are color-coded by state purity [Fig.~\ref{Fig:ST} (d)], the robust evolution remains confined to the sphere's surface, indicating a pure, unitary process, whereas the target-only path spirals inward---a clear signature of decoherence. 
These combined features demonstrate a capacity for universal noise resistance with minimal operational overhead.

This beneficial effect of achieving lower-amplitude and lower-fluence controls, as seen in Fig.~\ref{Fig:ST} (c), is a direct physical consequence of our optimization. Our total objective function $\mathcal{J}$ [Eq.~\eqref{Eq:J}] penalizes the noise sensitivity $D_{\rm eff}$. This metric $D_{\rm eff}$ is constructed from the kinetic coefficients $\gamma_{\alpha\alpha}[\omega_j(t)]$, which are related to the bath's spectral density $J(\omega)$ evaluated at the system's instantaneous Bohr frequencies $\omega_j(t)$. In our super-Ohmic model [$J(\omega) \propto \omega^3 \e^{-\omega/\omega_{\rm c}}$], this penalty is highly sensitive to large $\omega_j(t)$ values caused by ``aggressive" pulses (large amplitudes and/or rapid oscillations). Consequently, by minimizing $D_{\rm eff}$, our optimizer is explicitly and strongly penalized for using such aggressive controls. It is therefore guided to find a ``smoother" dynamical path to the target, which naturally corresponds to the gentler, lower-amplitude, and lower-fluence control fields observed in our results.

\newpage

\begin{figure*}[htp!]
\centering
\includegraphics[width=1.0\textwidth]{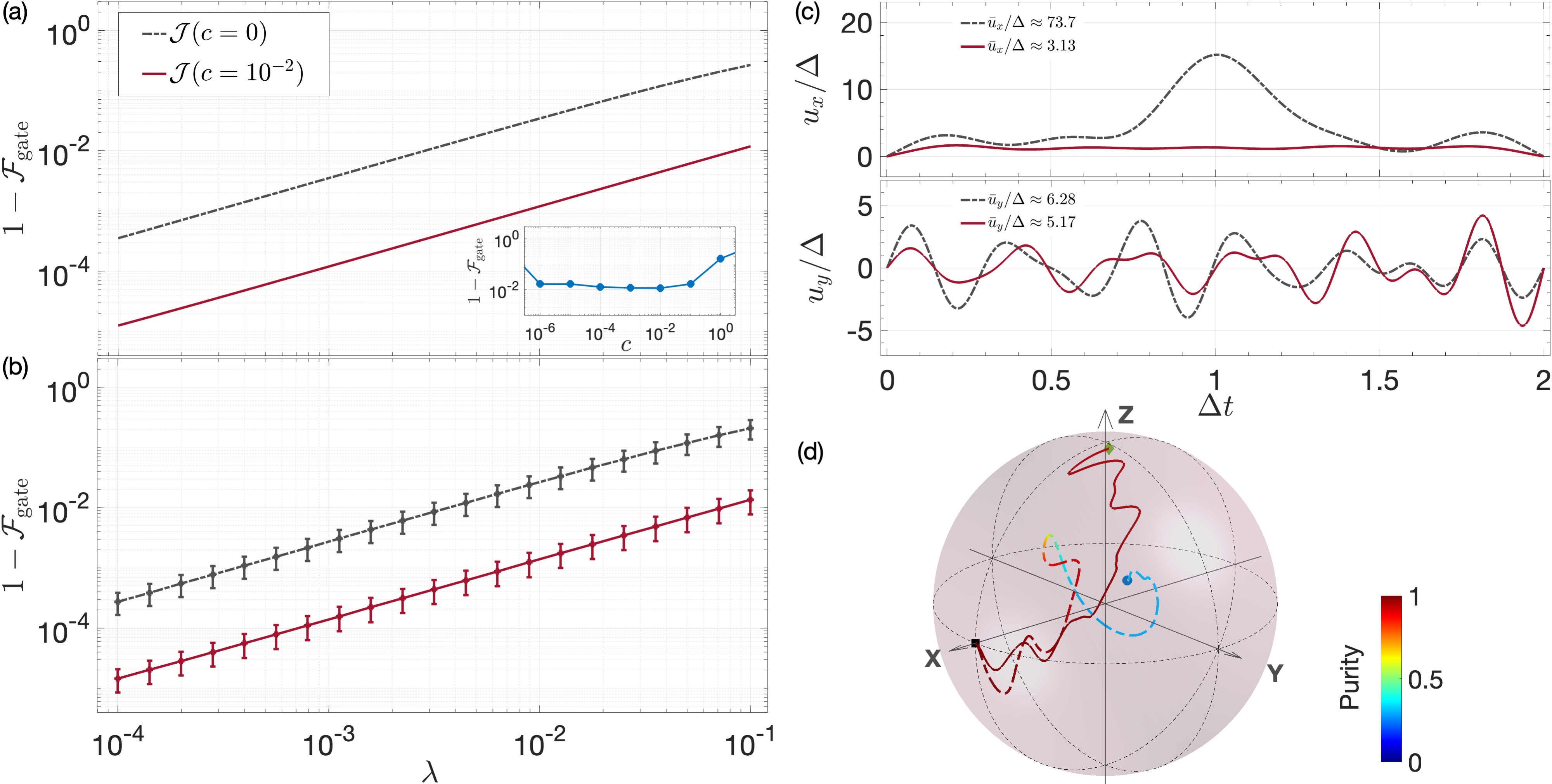}
\caption{
Universally robust control for single-qubit Hadamard gate.
Dash-dotted lines: target-only control ($c=0$); Solid lines: universally robust control ($c=10^{-2}$).
(a) Gate infidelity $1-\mathcal{F}_{\rm gate}$ versus coupling strength $\lambda$ for specific noise ($\hat{A} = \hat{\sigma}_z$).
Inset: Infidelity versus $c$ for $\lambda=0.1$.
(b) $1-\mathcal{F}_{\rm gate}$ for generic noise ($\hat{A} = \mathbf{n}\cdot\hat{\boldsymbol{\sigma}}$; $\mathbf{n}$: random unit vector).
Data averaged over 100 realizations and the error bars indicate the standard deviation.
(c) Optimized control fields $u_x(t)$ and $u_y(t)$ for $\hat{A} = \hat{\sigma}_z$ and $\lambda=0.1$.
The values of the time-integrated control power, $\bar{u}_{x,y}=\int_{0}^{\tau }u_{x,y}^{2}(t){\rm d}t$, are indicated in the legends.
(d) State evolution trajectories on Bloch sphere under the Hadamard gate (from the $x$ to the $z$ direction) for $\hat{A} = \hat{\sigma}_z$ and $\lambda=0.1$.
The evolution trajectory is color-coded to represent the purity of the state. Initial state: black square. Final states: blue dot (target-only) and green diamond (universally robust control). 
Parameters: Total control time $\tau=2/\Delta$ and inverse temperature $\beta = 1/\Delta$. 
}
\label{Fig:H}
\end{figure*}

\subsubsection{Quantum Gate Operations}

We extend our robust control protocol to the more demanding task of quantum gate synthesis, which requires noise-resilient state transformations across the entire Hilbert space---a significantly more complex challenge than single-state transfer \cite{Palao_2003_OC, Aspman_2024_Gate}.
We demonstrate protocol effectiveness for two fundamental gates: the single-qubit Hadamard gate and the two-qubit controlled-Z (CZ) gate, which form a universal gate set when combined with T gates \cite{nielsen_chuang_2010}.

Let's first consider the single-qubit Hadamard gate with target transformation
\begin{equation}
\hat{U}_{\rm H}=\frac{1}{\sqrt{2}}
\begin{pmatrix}
1 & 1 \\
1 & -1
\end{pmatrix}
\,.
\end{equation}
Using the control Hamiltonian from Eq. \eqref{Eq:H_ST}, we observe performance trends mirroring the two-level state transfer results.
The influence of the noise strength $\lambda $ on the gate fidelity with $\hat{A}=\hat{\sigma}_{z}$ and $\hat{A}= {\bf n}\cdot \hat{\boldsymbol \sigma}$
are depicted in Fig.~\ref{Fig:H} (a) and (b), respectively.
In both cases, the target-only optimization results deviate substantially from the ideal value once $\lambda \neq 0$, while those of robust control exhibit strong noise robustness (e.g., preserving $1-\mathcal{F}_{\text{gate}} < 0.02$ at $\lambda=0.1$ [Fig. \ref{Fig:H} (b)]).
Compared to the target-only optimization, the fidelity degradation of the robust protocol is suppressed by two orders of magnitude.
The comparison of the required control fields in the two protocols is shown in Fig.~\ref{Fig:H} (c), where we find again that the robust optimization requires much lower control-field amplitudes and fluences than the target-only one.
Besides, robust trajectories maintain confinement to the Bloch sphere surface (purity preservation), contrasting with target-only penetration (purity loss, as visually confirmed by the purity-coding of
the trajectory) [Fig. \ref{Fig:H} (d)].

\begin{figure*}[htp!]
\centering
\includegraphics[width=1.0\textwidth]{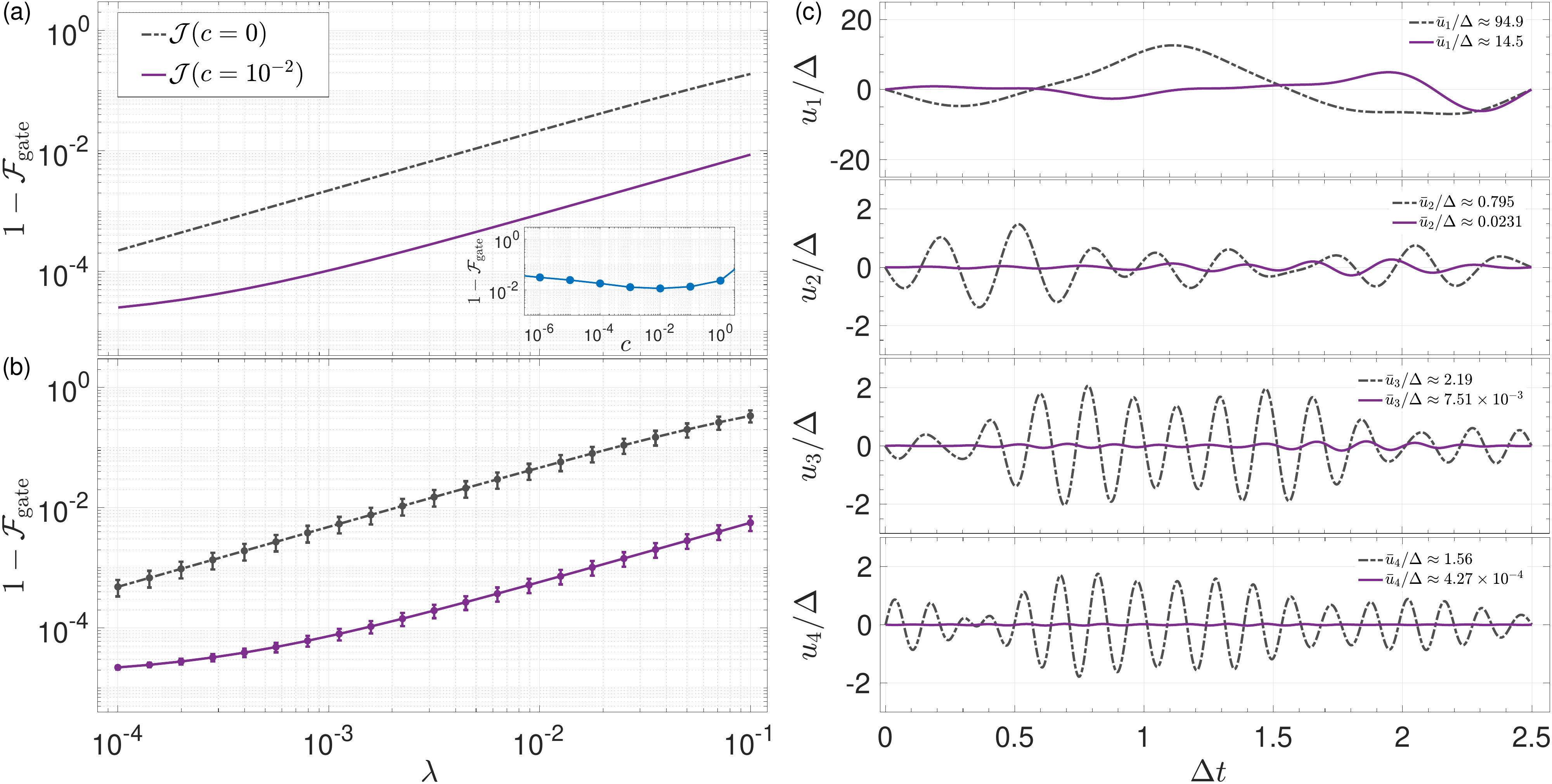}
\caption{
Universally robust control for two-qubit CZ gate.
Dash-dotted lines: target-only control ($c=0$); solid lines: universally robust control ($c=10^{-2}$).
(a) Gate infidelity $1-\mathcal{F}_{\rm gate}$ versus coupling strength $\lambda$ for specific noise ($\hat{A}=\hat{\sigma}_y  \otimes \hat{\sigma} _0$).
Inset: Infidelity versus $c$ for $\lambda=0.1$.
(b) $1-\mathcal{F}_{\rm gate}$ versus $\lambda$ for generic noise [$\hat{A}=\sum_{\mu\nu}a_{\mu\nu} \hat{\sigma} _\mu  \otimes \hat{\sigma} _\nu$, where the random coefficients $a_{\mu\nu}\sim N(0,1)$ with $\sum_{\mu\nu}a^2_{\mu\nu} = 1$].
Data averaged over 100 realizations and the error bars indicate the standard deviation. 
(c) Optimized control fields $u _{j}(t)$ ($j=1,2,3,4$) for $\hat{A}=\hat{\sigma} _y  \otimes \hat{\sigma} _0$ and $\lambda =0.1$.
The values of the time-integrated control power, $\bar{u}_{j}=\int_{0}^{\tau }u_{j}^{2}(t){\rm d}t$, are indicated in the legends. 
Parameters: Total control time $\tau=2.5/\Delta$ and inverse temperature $\beta = 1/\Delta$. 
}
\label{Fig:CZ}
\end{figure*}

For the two-qubit CZ gate, the target transformation is given by:
\begin{equation}
\hat{U}_{\rm CZ}=
\begin{pmatrix}
1 & 0 & 0 & 0 \\
0 & 1 & 0 & 0 \\
0 & 0 & 1 & 0 \\
0 & 0 & 0 & -1
\end{pmatrix}\,.
\end{equation}
The noise-free control Hamiltonian for this gate is organized as \cite{Riaz_2019_OC}
\begin{equation}
\begin{aligned}
\hat{H}_{\rm S}(t) = \Delta \hat{\sigma} _z \otimes \hat{\sigma} _z
&+ u_1(t)\hat{\sigma} _y  \otimes \hat{\sigma} _0 + u_2(t) \hat{\sigma} _0 \otimes \hat{\sigma} _y \\
&+ u_3(t) \hat{\sigma} _z  \otimes \hat{\sigma} _0 + u_4(t) \hat{\sigma} _0 \otimes \hat{\sigma} _z \, ,
\end{aligned}
\end{equation}
where $\hat{\sigma} _0$ is the identity operator for a qubit.
The control fields to be optimized are represented by $u _{j}(t)$ ($j=1,2,3,4$).
Fig.~\ref{Fig:CZ} shows the variation of the gate fidelities against $\lambda $ for (a) $\hat{A}=\hat{\sigma} _y  \otimes \hat{\sigma} _0$ and (b) $\hat{A}=\sum_{\mu\nu}a_{\mu\nu} \hat{\sigma} _\mu  \otimes \hat{\sigma} _\nu$.
Here, $\mu,\,\nu = 0,x,y,z$, and the random coefficients $a_{\mu\nu}\sim N(0,1)$ with $\sum_{\mu\nu}a^2_{\mu\nu} = 1$.
In both cases, we find qualitatively the same behaviors as those of the single-qubit gate.
That is, the infidelities of the target-only optimization increase rapidly as $\lambda $ increases, while the robust control can achieve and maintain high fidelity across a considerable range of noise amplitudes (e.g., preserving $1-\mathcal{F}_{\text{gate}} < 0.01$ at $\lambda=0.1$ [Fig. \ref{Fig:CZ} (b)]).
Besides, as shown in Fig.~\ref{Fig:CZ} (c), the robust control requires significantly lowered control-field amplitudes and fluences. This efficiency, also seen in the lower time-integrated power, is a key advantage for
scaling to multi-qubit systems where control power is a significant resource. 
The robust protocol consistently outperforms target-only optimization across all metrics, demonstrating scalability to multi-qubit systems while maintaining noise resilience and control efficiency.

\section{Discussion}

We have developed and validated a universally robust control framework capable of suppressing noise in open quantum systems without requiring specific knowledge of the noise channels.
By engineering a control-objective functional that minimizes an intrinsic noise-susceptibility metric, our method achieves robustness against arbitrary Markovian noise within the first-order weak coupling approximation.
Numerical simulations of state transfer and quantum gate operations show that this approach yields near-unity fidelities and suppresses errors by orders of magnitude compared to target-only optimization, even when subjected to diverse and poorly characterized noise sources.

This work represents a practical advancement in robust control design.
Traditionally, high-performance control has relied on methods like dynamical decoupling or filter-function engineering, which require at least partial characterization of the noise environment---a persistent experimental bottleneck.
Our framework circumvents this requirement, decoupling control design from the arduous task of environmental characterization.
This not only reduces experimental overhead but also provides resilience against uncharacterized or time-varying noise sources, a common challenge in real-world quantum devices. 
To make our results of direct experimental relevance, let us briefly discuss the connection between our simulation parameters and those in realistic experiments.
For current experiments on superconducting transmon qubits, $\Delta =10$ MHz appears to be practical \cite{Morten_2020_Review}, which allows us to translate the control time $\tau $, chosen in our simulations as 2/$\Delta $ for single-qubit operations and 2.5/$ \Delta $ for the two-qubit gate, to 200 ns and 250 ns, respectively.
These are well within the range of high-speed gate operations in current experiments \cite{Morten_2020_Review, Li_2023_npj, Simon_2023_nature}, and are orders-of-magnitude smaller than the typical coherence times found in experimental systems of either superconducting qubits (orders of milliseconds) \cite{Place_2021_nc, Somoroff_2023_prl, Tuokkola_2025_nc} or trapped ions (from seconds to minutes) \cite{Harty_2014_prl,Wang_2021_nc}. 
This vast separation of timescales powerfully demonstrates that our protocol is not only robust but also extremely fast, leaving a massive time window for performing complex algorithms before decoherence becomes a limiting factor. 

Looking forward, several avenues for future research are apparent. 
The current framework is derived under the Born-Markov approximation. Extending this approach to combat non-Markovian noise, which possesses memory effects, is a crucial next step and would broaden its applicability to an even wider range of physical systems.
While our method relies on a first-order expansion of the noise effects, its formulation allows for the inclusion of higher-order terms from Eq.
\eqref{Eq:Uerr}. Investigating the trade-off between the enhanced robustness from higher-order corrections and the increased computational cost of optimization would be a valuable pursuit.
Regarding the aspect of noise sources, apart from the environmental noise considered in this paper, practical experimental platforms may also suffer from coherent errors associated with the unitary part of the dynamics.
The origin of these errors can be traced back to system miscalibration \cite{Sheldon_2016_pra}, sample aging \cite{Yan_2020_arXiv}, or the slow drift of an experimental setup \cite{Proctor_2020_nc}.
While a full analysis of these effects is beyond the scope of the current paper, our method may offer some incidental benefits.
As we have found in Figs.~\ref{Fig:ST}-\ref{Fig:CZ} (c), the robust control solutions are characterized by lower amplitudes and smoother profiles, which are often less susceptible to hardware distortions and bandwidth limitations.
We therefore hypothesize that our protocol may also exhibit enhanced resilience to certain types of coherent errors, and we frame this as a promising avenue for future investigation.

Furthermore, the scalability of the optimization process presents a practical consideration.
Although we have demonstrated success for a two-qubit gate, the computational resources required to calculate and minimize $D_{\rm eff}$ will grow with system size.
Developing more efficient numerical techniques or leveraging machine learning to navigate the vast parameter space of many-qubit systems will be essential for applying this framework to larger-scale quantum processors.
In conclusion, our universally robust control protocol provides a powerful and practical tool for mitigating decoherence in quantum technologies.
Its versatile nature and independence from noise-model specification offer a flexible solution applicable to quantum computing \cite{Berberich_2024}, high-precision quantum sensing \cite{Hecht_2025_NC}, and other quantum control tasks plagued by poorly characterized noise.
By bridging a critical gap between theoretical control design and experimental reality, this work establishes a concrete and promising pathway toward the realization of high-fidelity quantum information processing.

\section{Methods}

\subsection{Master equation and kinetic coefficients}

To numerically implement the control protocol, it is convenient to rewrite
the master equation Eq.~\eqref{Eq:GKLS} in a more concrete form which
explicitly manifests the bath properties.
For the quantum tasks we considered
in the Results section, the system-bath interaction assumes the form $\hat{H}_{\rm I}=g\hat{A}\otimes \hat{B}$, and therefore the sum over the subscript $\alpha $ in Eq.~%
\eqref{Eq:kappa} disappears, yielding
\begin{equation}
\begin{aligned} \frac{\d}{\d t}\hat{\rho} _{\rm S}(t)= &- {\rm i} \left[\hat{H}_{\rm S}(t),\hat{\rho}_{\rm S}(t)\right] \\ &+g^{2}\sum_{j}\left[\eta _{j}
(t)\right]^{2}\gamma\left[\omega _{j}(t)\right] \left[\hat{F}_{j}(t)\hat{\rho}_{\rm S}(t)\hat{F}_{j}^{\dagger}(t)-\frac{1}{2}\{\hat{F}_{j}^{\dagger}(t)\hat{F}_{j}(t),\hat{\rho}_{\rm S}(t)\}\right]\,.
\end{aligned}
\label{Eq:rhoS_IV}
\end{equation}
Substituting the bath operators $\hat{H}_{\mathrm{B}}=\sum_{k}\left( \nu_{k}\hat{a}_{k}^{\dagger }\hat{a}_{k}+\frac{1}{2}\right) $ and $\hat{B} =\sum_{k}\left( g_{k}/g\right) (\hat{a}_{k}+\hat{a}_{k}^{\dagger })$ into the kinetic coefficients, we have [see details below Eq.~\eqref{Eq:kappa}]
\begin{eqnarray}
\gamma (\omega _{j} ) &=&\int_{-\infty }^{\infty }\mathrm{d}s\,\mathrm{e}^{-\mathrm{i}\omega _{j} s}\,\mathrm{Tr}_{\mathrm{B}}\left[\tilde{B}(s)\tilde{B}(0)\hat{\rho}_{\rm B}\right]  \nonumber \\
&=& \int_{-\infty }^{\infty }\mathrm{d}s\,\mathrm{e}^{-\mathrm{i}\omega _{j} s}\,\sum_{k}\frac{g_{k}^{2}}{g^{2}}[(\overline{N}\left( \nu _{k}\right) +1)\mathrm{e}^{-\mathrm{i}\nu _{k}s}+\overline{N}\left( \nu
_{k}\right) \mathrm{e}^{\mathrm{i}\nu _{k}s}]  \nonumber \\
&=&\sum_{k}\frac{g_{k}^{2}}{g^{2}}[2\pi (\overline{N}\left( \nu_{k}\right) +1)\delta (\omega _{j} +\nu _{k})+2\pi \overline{N}\left( \nu_{k}\right) \delta (\omega _{j} -\nu _{k})]  \, ,
\label{Eq:gamma}
\end{eqnarray}
where $\overline{N}\left( \nu _{k} \right) =1/\left[\exp(\beta\nu_k)-1\right]$ is the average occupation number given by the
Bose-Einstein statistics.
It is convenient to introduce the   spectral density function, $J(\tilde{\omega})=\sum_{k}(g_{k}/g)^{2}\delta (\tilde{\omega}-\nu _{k})$, by converting the summation into an integral, namely $%
\sum_{k}(g_{k}/g)^{2}\longrightarrow \int_{0}^{\infty }\mathrm{d}\tilde{\omega}\,J(\tilde{\omega})$. 
With this substitution, Eq.~\eqref{Eq:gamma} reduces to
\begin{equation}
\gamma (\omega _{j} )=2\pi \int_{0}^{\infty }\mathrm{d}\tilde{\omega}J(\tilde{\omega})[(\overline{N}\left( \tilde{\omega}\right) +1)\delta (\tilde{\omega} +\omega _{j} )+\overline{N}\left( \tilde{\omega}\right) \delta (\tilde{\omega} - \omega _{j} )].
\label{Eq:gamma2}
\end{equation}
The rates for absorption $(\omega _{j} >0)$ and emission $(\omega _{j} <0)$ can be easily read from Eq.~\eqref{Eq:gamma2} as $\gamma (\left\vert \omega _{j}
\right\vert )=2\pi J(\omega _{j} )\overline{N}\left( \omega _{j} \right) $ and $\gamma (-\left\vert \omega _{j} \right\vert )=2\pi J(\left\vert \omega _{j} \right\vert )(\overline{N}\left( \left\vert \omega _{j} \right\vert
\right) +1)$, respectively.

\subsection{Calculation of the Noise Sensitivity Metric $D_{\mathrm{eff}}$}

We now briefly discuss how to obtain the noise sensitivity metric $D_{\mathrm{eff}}$ in practical calculations.
The procedure is basically constituted by the following two steps:

1). The construction of the dynamical basis.
The calculation begins with the numerical solution of the time-dependent Schr\"{o}dinger equation for the system under a given control Hamiltonian $\hat{H}_{\rm S}(t)$. This yields the system propagator $\hat{U}_{\rm S}(t)$. At each time step of the numerical integration, $\hat{U}_{\rm S}(t)$ is diagonalized to determine the time-dependent eigenoperators (Lindblad jump operators) $\hat{F}_j(t)$ and the instantaneous Bohr frequencies $\omega_j(t)$. These form the natural, time-dependent basis in which the dissipative dynamics are described.

2). Numerical Integration and Norm Calculation. With the kinetic coefficients and constructed dynamical eigenoperators $\hat{F}_j(t)$, the operator $\mathbf{F}_{\alpha}$ in Eq.~\eqref{Eq:Deff} is calculated by numerically integrating the term $\sum_{j}\gamma_{\alpha\alpha}[\omega_{j}(t_{1})]\tilde{\mathbf{F}}^{j}(t_{1})$ over the total evolution 
time $\tau$. Here, $\tilde{\mathbf{F}}^{j}(t)$ is the vectorized jump superoperator in the interaction picture.
Finally, the value of $D_{\text{eff}}$ is computed from Eq.~\eqref{Eq:Deff}.

The above two steps serve as a practical recipe for implementing our method.

\section*{DATA AVAILABILITY }

Data is available from the corresponding author upon reasonable request.

\section*{CODE AVAILABILITY  }

The codes used to generate data for this paper are available from the corresponding author upon reasonable request.

\bibliography{Reference}

\section*{Acknowledgments}
X.Q. acknowledges support from Shanghai Science and Technology project (24LZ1401600) and the Fundamental Research Funds for the Central Universities.
J.F. acknowledges support from the National Key R\&D Program
of China under Grant No. 2022YFA1404201, the National
Natural Science Foundation of China (NSFC) under Grants
No. 12174233, the Research Project Supported by
Shanxi Scholarship Council of China and Shanxi `1331KSC'.

\section*{AUTHOR CONTRIBUTIONS}
X.Q. initiated the work and designed the proposed architecture with feedback from all authors. L.D. and X.Q.
developed the methods and performed numerical calculations. J.F. and X.Q. supervised the project. All authors contributed in completing the paper.

\section*{COMPETING INTERESTS}
The authors declare no competing interests.

\end{document}